\documentstyle[12pt,epsf]{article}
\renewcommand{\thefootnote}{\fnsymbol{footnote}}
\newlength{\extraspace}
\newlength{\extraspaces}
\newcommand{\be}{\begin{equation}
\addtolength{\abovedisplayskip}{\extraspaces}
\addtolength{\belowdisplayskip}{\extraspaces}
\addtolength{\abovedisplayshortskip}{\extraspace}
\addtolength{\belowdisplayshortskip}{\extraspace}}
\newcommand{\ee}{\end{equation}}
 
\newcommand{\ba}{\begin{eqnarray}
\addtolength{\abovedisplayskip}{\extraspaces}
\addtolength{\belowdisplayskip}{\extraspaces}
\addtolength{\abovedisplayshortskip}{\extraspace}
\addtolength{\belowdisplayshortskip}{\extraspace}}
\newcommand{\ea}{\end{eqnarray}}

\newcommand{\bas}{\begin{eqnarray*}
\addtolength{\abovedisplayskip}{\extraspaces}
\addtolength{\belowdisplayskip}{\extraspaces}
\addtolength{\abovedisplayshortskip}{\extraspace}
\addtolength{\belowdisplayshortskip}{\extraspace}}
\newcommand{\eas}{\end{eqnarray*}}
 
\newcounter{subequation}[equation]
\makeatletter

\expandafter\let\expandafter
\reset@font\csname reset@font\endcsname

\def\subeqnarray{\arraycolsep1pt
    \def\@eqnnum\stepcounter##1{\stepcounter{subequation}%
        {\reset@font\rm(\theequation\alph{subequation})}}
\jot5mm     \eqnarray}

\makeatother

\newcommand{\NP}[1]{Nucl.\ Phys.\ {\bf #1}}

\newcommand{\CMP}[1]{Comm.\ Math.\ Phys.\ {\bf #1}}

\newcommand{\C}{\mbox{$\,${\sf I}\hspace{-1.2ex}{\bf C}}}
\newcommand{\Z}{\mbox{{\sf Z}\hspace{-1ex}{\sf Z}}}
\newcommand{\R}{\mbox{\rm I\hspace{-.4ex}R}}
\newcommand{\1}{\mbox{1\hspace{-.8ex}1}}
\newcommand{\bra}{\langle}
\newcommand{\ket}{\rangle}
\newcommand{\ra}{\rightarrow}

\newcommand{\rra}{\ \longrightarrow \ }

\newcommand{\is}{ &\!=\!& }
\newcommand{\nonum}{\nonumber \\[1.5mm]}
\newcommand{\sspace}{\makebox[1cm]{ }}

\newcommand{\nspace}{\!\!\!\!\!\!\!\!\!\!}

\newcommand{\th}{{\theta}}
\newcommand{\eps}{{\epsilon}}
\newcommand{\lb}{\lambda}
\newcommand{\sh}{{\rm sh}}
\newcommand{\ch}{{\rm ch}}

\newcommand{\cP}{{\cal P}}

\newcommand{\dd}{\partial}

\newcommand{\om}{\omega}

\newcommand{\ben}{\begin{displaymath}}
\newcommand{\een}{\end{displaymath}}
\newcommand{\he}{\widehat{{\bf e}}}
\newcommand{\hf}{\widehat{{\bf f}}}
\newcommand{\hh}{\widehat{{\bf h}}}

\begin{document}
\begin{titlepage}
%
\renewcommand{\thefootnote}{\fnsymbol{footnote}}
\mbox{} 
\vspace{15mm}
 
\begin{center}
{\LARGE Quantized Einstein-Rosen waves, ${\rm AdS}_2$,}\\[3mm] 
{\LARGE and spontaneous symmetry breaking}\\[2.5cm]

{\large Max Niedermaier\footnote{E-mail: nie@prospero.phyast.pitt.edu}}\\ [3mm]
{\small\sl Department of Physics} \\
{\small\sl 100 Allen Hall, University of Pittsburgh} \\
{\small\sl Pittsburgh, PA 15260, USA}
\vspace{2.2cm}

{\bf Abstract}
\end{center}
\begin{quote}
4D cylindrical gravitational waves with aligned polarizations (Einstein-Rosen 
waves) are shown to be described by a weight $1/2$ massive free field on the 
double cover of ${\rm AdS}_2$. Thorn's C-energy is one of the $sl(2,\R)$ 
generators, the reconstruction from the (timelike) symmetry axis is the 
${\rm CFT}_1$ holography. Classically the phase space is also invariant under 
a O(1,1) group action on the metric coefficients that is a remnant of the 
original 4D diffeomorphism invariance. In the quantum theory this symmetry
is found to be spontaneously broken while the ${\rm AdS}_2$ conformal invariance 
remains intact. 
\end{quote}
\vfill

\renewcommand{\thefootnote}{\arabic{footnote}}
\setcounter{footnote}{0}
\end{titlepage}

{\em Introduction:} The Einstein-Rosen (ER) subsector of general relativity 
provides a simple, yet instructive, laboratory for studying certain quantum 
aspects of gravity \cite{Kuchar,AshP,Ash}. In brief, ER-waves are 
gravitational wave solutions to the Einstein equations with cylindrical 
symmetry and aligned polarizations. In Weyl-canonical coordinates the 
4-dim.~line element can locally be written in the form
\be
d s^2 = e^{\gamma - \phi}(dt^2 - dr^2) - (e^{\phi}dz^2 + r^2 
e^{-\phi} d\alpha^2)\;,
\label{ER1}
\ee
where $\phi$ and $\gamma$ are functions of $t,r$ and $z,\alpha$ are 
the coordinates along the orbits of the two Killing vector fields.
Einstein's equations turn out simply to be equivalent to a spherical
wave equation for $\phi$, i.e.~$[\dd_t^2 - \dd_r^2 - \frac{1}{r}
\dd_r]\phi =0$, together with the condition that $\gamma$ is expressed 
in terms of $\phi$ by 
\be
\gamma(t,r') = \frac{1}{2}\int_0^{r'} dr r[(\dd_t\phi)^2 + (\dd_r \phi)^2]\;,
\label{ER3}
\ee  
The conserved quantity $\gamma_{\infty} = \gamma(t, \infty)$ is 
known as Thorn's C-energy while the physical Hamiltonian is 
$1- e^{-\gamma_{\infty}/2}$ \cite{AshV}, and measures a 3D deficit angle.

The Killing vector fields $\frac{\partial}{\partial z}$, 
$\frac{\partial}{\partial \alpha}$ of course are unique only up
to normalization. A constant rescaling of the coordinates 
$z \ra e^{-\lb/2} z,\; \alpha \ra e^{\lb/2} \alpha$, $\lb \in \R$, amounts 
to changing their norm, which can be compensated by $\phi(t,r) \ra \phi(t,r) + \lb$. 
For the combinations of the metric coefficients $\ch \phi$
and $\sh \phi$ the shift amounts to a linear transformation by an element of the 
non-compact Lie group O(1,1), which can be seen to be a symmetry of 
the classical phase space. Adopting the quantization scheme of \cite{AshP}
we shall later find this symmetry to be spontaneously 
broken in the quantum theory.

In the $(t,r)$ part of the line element (\ref{ER1}) the shift $\phi(t,r) \ra 
\phi(t,r) + \lb$ can be compensated by a rescaling $t \ra e^{\lb/2} t,\,
r \ra e^{\lb/2} r$. Together with the time translations this generates a Borel 
subgroup of $SL(2,\R)$, which classically can be promoted to a symmetry of 
(\ref{ER1}). Although the 
``targetspace'' O(1,1) symmetry is spontaneously broken in the quantum theory,
the dilatation symmetry $t \ra e^{\lb/2} t,\,r \ra e^{\lb/2} r$ turns out to remain 
intact, and is unitarily implemented by a generator $H$. In particular 
this gives rise to a thermalization phenomenon akin to the the Unruh effect: 
Restricting the quantum theory to the cone $t \geq r\geq 0$, the ground state 
for $\gamma_{\infty}$ (generating the $t$-evolution) is a thermal state for the 
dilatation-evolution, of temperature $1/2\pi$. From the 4D viewpoint the restricted 
theory can be regarded as the quantum theory of those ``exotic'' ER-waves whose 
scalar $\phi(t,r)$ has support for $t \geq r \geq 0$ only. On the level of the 
Lie algebra, the symmetry can further be extended to a full $sl(2,\R)$ action 
on the $(t,r)$ ``worldsheet'', whose generators $E = \gamma_{\infty}/2,\,H$ and 
$F$ are local conserved charges that leave the vacuum invariant. The corresponding 
finite transformations however will be symmetries of the (classical and quantum) 
theory only if the $(t,r)$ Lorentzian space is extended to the double cover of 
two-dimensional Anti-deSitter space, ${\rm AdS}_2$. 
Without the extension the ER-system would also not allow for a CPT operation. 
Taking advantage of the ${\rm AdS}_2$ extension, the ${\rm AdS}_2/{\rm CFT}_1$ 
correspondence then yields a quantum version of the known classical 
reconstruction \cite{trec} from the $r=0$ symmetry axis.

{\em Spontaneous breakdown of O(1,1) symmetry:} 
The quantum theory of ER-waves descends from that of a free scalar field 
$\Phi = \sqrt{r}\phi$ having the following expansion in terms of Bessel 
functions (see e.g.~\cite{AshP})
\ba
&& \Phi(t,r) = \int_0^{\infty} d\om \sqrt{\frac{r}{2}} 
J_0(\om r)\left[A(\om) e^{-i\om t} +
A^{\dagger}(\om)e^{i\om t}\right] \;,
\nonum
&& [A(\om),A^{\dagger}(\om')] = \delta(\om-\om')\;.
\label{Phidef1}
\ea
It is readily identified as that of a weight $h=1/2$ massive free
field on ${\rm AdS}_2$. Indeed in Poincar\'{e} coordinates $(t,r)$ the 
wave equation on ${\rm AdS}_2$ is (see e.g.~\cite{Strom,Bertetal}) 
\be
\big\{ \Box_{AdS_2} + h(h-1) \big\} \varphi(t,r) =0\;,\sspace
\Box_{AdS_2} = r^2(\dd_t^2 - \dd_r^2) \;,
\label{AdSwave}
\ee
where $h\geq 1/2$ parameterizes the mass via $m^2 = h(h-1)$, and $h=1/2$ is 
the unitarity threshold. For $h=1/2$ one obtains the expansion (\ref{Phidef1}) 
in terms of positive and negative frequency solutions. The shift invariance
$\phi(t,r) \ra \phi(t,r) + \lb$ reflects the ambiguity in the 
zero mode contribution proportional to $\sqrt{r}$ in (\ref{Phidef1}). 
The $(t,r)$ coordinates cover only part of ${\rm AdS}_2$; we shall 
describe later why the extension to the double cover of ${\rm AdS}_2$ 
is mandatory in this context. 

For the discussion of symmetry breaking it is more useful to regard
(\ref{Phidef1}) as the spherical reduction of a 1+2 dim.~massless scalar field.
Recalling the positive frequency two-point function of the latter
\be
W_3(x) = \frac{1}{4\pi}\left[\th(x^2)\frac{{\rm sign(x_0)}}{\sqrt{x^2}}
+ i \th(-x^2) \frac{1}{\sqrt{-x^2}}\right]\,,\quad x^2 = x_0^2 - x_1^2 - x_2^2\,,
\label{3d}
\ee
and switching to angular coordinates in the $x_1,x_2$ plane, the two-point
function of $\Phi(t,r)$ in the Fock vacuum $A(\omega) |0\ket =0$, can be 
written as
\ba
W(t_1-t_2;r_1,r_2) \is i \bra 0|\Phi(t_1,r_1)\,\Phi(t_2,r_2) |0\ket =
\sqrt{r_1 r_2} \int_{-\pi}^\pi d\alpha \,W_3(t_1-t_2;r_1,r_2;\alpha)\;,
\label{2point}
\ea
in an obvious notation for the integrand.
The integrations in (\ref{2point}) can be reduced to complete elliptic integrals 
${\bf K}(u) = \int_0^{\pi/2}d \alpha (1 - u \sin^2\alpha)^{-1/2}$. The result is 
\be
W(t_1-t_2;r_1,r_2) = \frac{1}{2}{\rm sign}(t_1-t_2)\,d(\xi^2) 
+ \frac{i}{2} \,d(1 -\xi^2)\;,
\label{D1}
\ee
where $\xi^2 = [(t_1-t_2)^2 - (r_1 - r_2)^2]/(4 r_1 r_2)$ 
is the ${\rm AdS}_2$ invariant distance in Poincar\'{e} coordinates.
The function $d(u)$ is given by $d(u) =0$ for $u <0$ and  
\be
d(u) = \th(1-u) \frac{1}{\pi} \,{\bf K}(u) + \th(u-1)
\frac{1}{\pi \sqrt{u}}\,{\bf K}(1/u)\,,\quad u \geq 0\;.
\label{D3}
\ee
Up to a sign $d(\xi^2)$ coincides with the commutator 
function $2 {\rm Re}\,W$; a plot of $d(\xi^2)$ is shown below. Large and small 
distances are related by the duality $d(1/\xi^2) = \sqrt{\xi^2}\, d(\xi^2)$.
The limits are: $\lim_{\xi^2 \ra 0^+} d(\xi^2) = 1/2$ and $d(\xi^2) \sim
1/2\sqrt{\xi^2}$, for $\xi^2 \ra \infty$. Note that rather than being singular 
on the 1+1 dim.~lightcone $\xi^2 =0$, $d(\xi^2)$ is singular at $\xi^2=1$, 
i.e.~$|t_1-t_2| = r_1 + r_2$. The behavior across the $1+1$ dim. lightcone is 
discontinuous, but with a finite jump.
\vspace{-13mm}

\begin{figure}[hbt]
\hspace{3.6cm}
\leavevmode
\epsfxsize=100mm
\epsfysize=65mm
\epsfbox{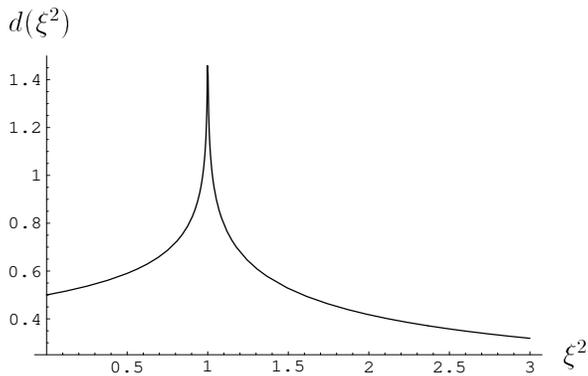}
\caption{Commutator function of the ER-scalar}
\end{figure}

Classically the symmetry $\Phi(t,r)  \ra \Phi(t,r)  + \lb \sqrt{r}$ is
generated by the conserved charge $Q = i\int_0^{\infty} dr \sqrt{r} \dd_t \Phi$.
In the quantum theory, following the standard procedure \cite{Swieca}, one 
will try to define $Q$ as a suitable limit of regularized operators $Q_R$
(supported in a sphere of radius $R$), as $R \ra \infty$. The symmetry associated 
with the current is said to be spontaneously broken, if the limit of the 
commutator's vacuum expectation value $q := \lim_{R \ra \infty}
(\Omega, [Q_R, A(x)] \Omega)\;(*)$ is a non-zero number, independent of the 
regulators, for some $A(x)$. In this case the limit of the operators $Q_R$ does 
not define the generator of a unitary group of automorphisms which induces the 
symmetry transformations and leaves the vacuum invariant. For the application 
to the ER-waves it is important that this criterion has a functional analytical 
origin \cite{Swieca} and does not hinge e.g.~on Poincar\'{e} invariance. In a 
Minkowski space quantum field theory it is known that $q \neq 0$ can only occur 
in the presence of massless 1-particle states. Indeed the prototype example 
for $(*)$ to hold with $q \neq 0$ is the current $i \dd_0 \phi$ of a massless 
free scalar field in $d \geq 3$ spacetime dimensions, taking for $A(x)$ the 
scalar field $\phi(x)$ itself. The symmetry broken is $\phi(x) \ra \phi(x) + \lb$. 
In $d=2$ this result does not apply because the massless commutator function 
fails to comply with the Wightman axioms, -- which is one way of understanding 
the absence of spontaneous symmetry breaking in two dimensions \cite{Coleman}. 

In view of (\ref{3d}), (\ref{2point}) the ER-scalar lies in-between 
the 2+1 dim.~and the 1+1 dim.~situation and the issue has to be examined from 
scratch. To do so, one first verifies that for every operator $A$ in the polynomial 
field algebra of $\Phi$ the commutator $[Q_R,A]$ is independent of $g_R$ 
and $f$ for sufficiently large $R$, where 
\be
Q_R = i \int dt dr \sqrt{r} f(t) g_R(r)\,\dd_t \Phi(t,r)\;,
\label{Qdef}
\ee 
and $g_R$ is a smooth function with $g_R(r) =1$ for $r \in [R^{-1}, R]$, 
rapidly decaying for $r \ra 0$ and $r \ra \infty$, and $\int dt f(t) =1$. 
The independence of $g_R$ follows from the vanishing of the commutator
function $2 {\rm Re}W$ at spacelike 1+1-dim. distances; the independence from $f$
is a consequence of current conservation. The attempt 
to define a hermitian charge operator along the above lines requires that 
the commutator with $\Phi$ has vanishing vacuum expectation value for 
sufficiently large $R$. However one straightforwardly verifies
\be
\bra 0| [Q_R, \Phi(t_2,r_2) ]|0\ket = \sqrt{r_2} \;,\quad \mbox{for $R$ large}\;, 
\ee
establishing that the transformation $\Phi(t,r) \ra \Phi(t,r) + \lb \sqrt{r}$ 
cannot be unitarily implemented on the Fock space. It is crucial here
that the commutator function is a well-defined ${\rm AdS}_2$-invariant
distribution so that no caveat arises as in the case of a 
relativistic massless field in 1+1 dim. Minkowski space; c.f.~\cite{Coleman}.   
An analogous symmetry breaking has already been found in the situation 
with generic polarizations \cite{qernst} by very different means.

{\em Unbroken $sl(2,\R)$ symmetry:} 
The field eq.~(\ref{AdSwave}) follows from an obvious action, which
can be verified to be separately invariant under the variations $\delta_{e}\Phi 
= -i {\bf e} \Phi$, $\delta_{h}\Phi = -i {\bf h} \Phi$,
$\delta_{f}\Phi = -i {\bf f} \Phi$. Here ${\bf e},{\bf h}, {\bf f}$ 
generate a realization of $sl(2,\R)$ in terms of differential operators
${\bf e} = \dd_t,\,{\bf h} = 2(t\dd_t + r \dd_r),\,
{\bf f} = - (t^2 + r^2) \dd_t - 2 t r \dd_r$,
acting on a suitable space of test functions. In particular they are 
anti-hermitian wrt the measure $\int dt dt/r^2$. The quadratic Casimir is 
${\bf C} = -\Box_{AdS_2}$, so that the operator equation of motion for 
$\Phi$ amounts to ${\bf C} = h(h-1)|_{h = 1/2}\1 = -1/4 \1$.
The Noether currents $J_{\mu}({\bf e}), J_{\mu}({\bf h}), J_{\mu}({\bf f})$ 
associated with the above $sl(2,\R)$ variations are 
\ba
&\nspace \!\!&\begin{array}{ll}
J_t({\bf e}) = (\dd_t \Phi)^2 + (\dd_r \Phi)^2 - \frac{1}{4 r^2} \Phi^2 \;,
\;\; & J_r({\bf e}) = 2 \dd_r \Phi \dd_t \Phi\;,\\[2mm]
J_t({\bf h}) = 2 t J_t({\bf e}) + 2 r J_r({\bf e})\;,\;\; &
J_r({\bf h}) =  2 r[J_t({\bf e}) + \frac{1}{2 r^2} \Phi^2] + 
2 t J_r({\bf e})\;,\\[2mm]
J_t({\bf f}) = - (t^2 +r^2) J_t({\bf e}) - 2 t r J_r({\bf e})\;,\;\; &
J_r({\bf f}) = - 2 t r[J_t({\bf e}) + \frac{1}{2 r^2} \Phi^2] 
-(t^2 +r^2)J_r({\bf e}) \,.
\end{array}
\label{ERcurr}
\ea
In (\ref{ERcurr}) we omitted total derivative terms in order to 
simplify the expressions. Restoring them is however essential to arrive at the 
correct classical Noether charges. In the quantum theory normal ordering is 
understood. Repeating the previous analysis for these charges, one finds 
that this symmetry is unbroken and is implemented by hermitian operators 
$E,H,F$ that annihilate the vacuum. They satisfy 
$[E ,\Phi(t,r)] = -i {\bf e} \Phi(t,r)$, etc., and generate the expected 
$sl(2,\R)$ algebra: $[H,E] = - 2 i E,\,[F, E] = i H,\,[H,F] = 2 i F$.
Explicitly 
\ba
\begin{array}{ll}
\displaystyle{ E = \int_0^{\infty} d\omega A^{\dagger}(\omega)\, 
i e(\omega) A(\omega)} \;, \quad & e(\omega) = -i \omega\;,\\[2mm]
\displaystyle{H = \int_0^{\infty} d\omega A^{\dagger}(\omega)\, 
i h(\omega) A(\omega)} \;, \quad & h(\omega) = -(2 \omega \dd_{\omega} + 1)\;,
\\[2mm]
\displaystyle{F = \int_0^{\infty} d \omega A^{\dagger}(\omega)\, 
i f(\omega) A(\omega)} \;, \quad &
\,f(\omega) = -i(\omega \dd^2_{\omega} + \dd_{\omega})\;.
\end{array}
\label{qsl2}
\ea
Here $e(\omega),\,h(\omega),\,f(\omega)$ again form an anti-hermitian 
realization of $sl(2,\R)$, now with constant Casimir $C(\omega) = -1/4$.  
In particular $2 E$ yields a quantum version of the C-energy $\gamma_{\infty}$ 
\cite{AshP}. Exponentiating this $sl(2,\R)$ action yields 
\ba
&& \begin{array}{lll}
\displaystyle{e^{i s E} A^{\dagger}(\omega) e^{-i s E}} &\!\!=
\displaystyle{e^{-s e(\omega)} A^{\dagger}(\omega)} &\!\! = 
\displaystyle{e^{i s \omega} A^{\dagger}(\omega)}\;,
\\[2.5mm]
\displaystyle{ e^{i s H} A^{\dagger}(\omega) e^{-i s H}} & \!\!=
\displaystyle{ e^{s h(\omega)} A^{\dagger}(\omega)} & \!\!= 
\displaystyle{ e^{-s} A^{\dagger}(e^{-2 s} \omega)}\;,
\\[2mm]
\displaystyle{e^{i s F} A^{\dagger}(\omega) e^{-i s F}} & \!\!= 
\displaystyle{ e^{-s f(\omega)} A^{\dagger}(\omega)} & \!\!= 
\displaystyle{ -\frac{i}{s} \int_0^{\infty} d \lb\, 
J_0\left(\frac{2}{s} \sqrt{\omega \lb}\right)\! e^{\frac{i}{s}(\omega + \lb)}
A^{\dagger}(\lb)}\;. 
\end{array}
\label{qSL2}
\ea 
Together the 1-parameter families (\ref{qSL2}) generate a unitary representation 
$SL(2,\R) \ni g \ra U(g)$ on the Fock space. The 1-particle subspace of the Fock 
space is irreducible wrt this $SL(2,\R)$ action; see e.g.~\cite{VK}. 
The action of $U(g)$ is best described in terms of the Laplace transforms 
\be 
\!\!\!
A^-(\th) = \frac{1}{\sqrt{2}}\int_0^{\infty} \!d\omega A(\omega) e^{-i \omega \th},
\;{\rm Im}\,\th < 0\;,\quad 
A^+(\th) = \frac{1}{\sqrt{2}}\int_0^{\infty} \!d\omega 
A^{\dagger}(\omega) e^{i \omega \th},
\;{\rm Im}\,\th > 0\,,
\label{AFT}
\ee
(both supposed to vanish in the other half plane) and reads
\be
U(g) A^{\pm}(\th) U(g)^{-1} = \frac{A^{\pm}(\th^g)}{c \th + d}\;,
\sspace \th^g = \frac{ a \th +b}{c \th + d} \;,\quad
g= { a \;\; b \choose c\;\; d} \in SL(2,\R)\;.
\label{SL2AFT}
\ee 

We take $\phi^b(t) = \lim_{\eps \ra 0}\,[ A^-(t - i\eps) + A^+(t + i\eps)]$ 
as the definition of the (classical and quantum) field on the $r=0$ symmetry axis. 
In a distributional sense $\phi^b(t)$ inherits the transformation law 
(\ref{SL2AFT}). The two-point function on the axis 
\be
\lim_{r \ra 0}\frac{1}{r}W(t_1 -t_2;r,r) = \frac{1}{2(t_1 -t_2)}=
i\,\bra 0| \phi^b(t_1) \phi^b(t_2) |0\ket\;,
\ee
is well-defined and covariant.
The boundary field $\phi^b(t)$ is of additional interest because 
$\phi(t,r)$ can be uniquely reconstructed from it. Indeed, using 
the Laplace transforms (\ref{AFT}) one can rewrite the expansion
(\ref{Phidef1}) as 
\ba
\phi(t,r) \is  \frac{1}{2\pi}\int_{t-r}^{t+r} d \th 
\frac{1}{\sqrt{r^2 - (t - \th)^2}}\,[A^+(\th) + A^-(\th)]\nonum
&+&  \frac{i}{2\pi}\left( \int_{-\infty}^{t-r} d \th
- \int_{t+r}^{\infty} d\th \right)
\frac{1}{\sqrt{(t - \th)^2-r^2}}\,[A^+(\th) - A^-(\th)]\;.
\label{reconstr1}
\ea  
The terms in square brackets can be identified as the field on the axis
and its Hilbert transform 
\be
A^+(\th)+ A^-(\th) = \phi^b(\th) \;,\sspace 
A^+(\th)- A^-(\th) = \frac{1}{i\pi}\, \cP\!\!\int_{-\infty}^{\infty} 
dt\, \frac{\phi^b(t)}{t -\th}\;.
\ee
Thus (\ref{reconstr1}) allows one to reconstruct the field entirely 
from its values on the $t$-axis. In contrast to the 
Cauchy problem only one function has to be prescribed.

The finite transformations generated by the exponentials of 
${\bf e},\,{\bf h},\,{\bf f}$ are Moebius transformations in the null 
coordinates $t\pm r$, i.e.
\be
t\pm r \rra (t\pm r)^g = \frac{a(t\pm r) + b}{c(t \pm r) + d} \;,\quad ad - bc =1\;.
\label{SL2}
\ee
They leave the line element $ds^2 = (dt^2-dr^2)/r^2$ and the Casimir 
operator in (\ref{AdSwave}) invariant. Thus, although for $c \neq 0$ they no longer 
preserve the Weyl-canonical form of the line element (\ref{ER1}), they
in principle map solutions of the field equations into themselves.    
However outside some save regions (like the $t$-axis or the cone $t \geq r \geq 0$ 
for $c d > 0$) transformations with $c\neq 0$ can map positive $r$ and positive 
$t^2 - r^2$ into negative ones, and can also ruin the hermiticity of the 
transformed field $\phi$. Such problems are known to be a generic feature of 
conformal quantum field theories and the resolution is to switch to a suitable 
covering manifold (see e.g.~\cite{LM} in the original, and \cite{Bertetal} in the 
AdS context). In the case at hand one is lead to the double cover of ${\rm AdS}_2$.

{\em ${\rm AdS}_2$ and $SU(1,1)$ action:}
The universal cover of ${\rm AdS}_2$ can be described in terms of coordinates 
$(\tau,\sigma)$ in which the line element takes the form
$d s^2 = (d\tau^2 - d\sigma^2)/\cos^2 \sigma$, with 
$-\frac{\pi}{2} \leq \sigma < \frac{\pi}{2}$, $-\infty < \tau < \infty$. 
Spatial infinity is at $\sigma = \pm \frac{\pi}{2}$, 
the segment $-\pi < \tau < \pi$ is called
the primitive domain. The invariant distance is 
\be
\xi_G^2 = \frac{\cos(\sigma_1-\sigma_2) 
- \cos(\tau_1 -\tau_2)}{2 \cos\sigma_1 \cos\sigma_2}\;.
\ee
The projection onto ${\rm AdS}_2$ is effected by 
identifying the points $(\tau,\sigma)$ and $(\tau + \pi,-\sigma)$. 
The previously used Poincar\'{e} coordinates $(t,r)$ cover only the
patch $ \tau + \sigma < \frac{\pi}{2}$ and $\tau - \sigma > 
-\frac{\pi}{2}$ in the primitive domain; c.f.~Fig.~2 below. 
They are related to the global 
coordinates by $(t,r) = (\sin\tau,\cos\sigma)/(\cos\tau -\sin\sigma)$.
In brief, the problem with the special conformal transformations, $c\neq 0$, 
now is absent, because by virtue of 
$-r(\tau,\sigma) = r(\tau + \pi,-\sigma)$, a negative $r$ in the primitive
domain can always be traded for a positive $r$ in a secondary domain.

The ${\rm AdS}_2$ Laplacian in global coordinates is $\cos^2\sigma
(\dd_{\tau}^2 -\dd_{\sigma}^2)$. A complete set of positive frequency solutions 
of (\ref{AdSwave}) of weight $1/2$ is now provided by (e.g.~\cite{NY})  
\be
\varphi_n(\tau,\sigma) = \frac{1}{\sqrt{2}} e^{-i(n+1/2)\tau}(\cos\sigma)^{1/2}
P_n(\sin \sigma),\,\quad n\geq 0,
\label{phin}
\ee
where $P_n$ are the Legendre polynomials. In contrast to the 
ones in (2) they are square integrable in addition to being orthonormal wrt
the Klein-Gordon symplectic form. The $h=1/2$ free scalar field admits an 
alternative expansion in terms of the solutions (\ref{phin})
\be
\Phi(\tau,\sigma) = \sum_{n \geq 0} [a_n \,\varphi_n(\tau,\sigma) + 
a_n^{\dagger}\, \varphi_n^*(\tau,\sigma) ]\;,\sspace
[a_n,a^{\dagger}_m] = \delta_{mn}\;.
\ee
The shift invariance here corresponds to shifts by the two real linear combinations 
of the zero modes $v_+ = \sqrt{2}\cos\frac{\tau}{2} (\cos \sigma)^{1/2}$, 
$v_- = -\sqrt{2}\sin\frac{\tau}{2} (\cos \sigma)^{1/2}$. Infinitesimally they are 
generated by the hermitian operators $Q_+ = -i(a_0 - a_0^{\dagger})$ 
and $Q_- = a_0 +a_0^{\dagger}$, i.e.~$[Q_{\pm}, \Phi] = -i v_{\pm}$,
which are linked by  the Hamiltonian $\widehat{H}/2$ of the $\tau$-evolution via 
$[\widehat{H}, Q_{\pm}] = \pm i Q_{\mp}$. Clearly $Q_{\pm}$ don't leave the 
Fock vacuum invariant, consistent with the symmetry breaking found in the $sl(2,\R)$ 
framework. In fact no normalizable state exists annihilated by both $Q_+$ and $Q_-$. 

The Fock vacuum in (\ref{2point}) is also the Fock vacuum for the $a_n$-modes
\cite{Strom}. The global two-point function in the complex $(\tau,\sigma)$ 
domain is 
\be
W(\tau_1 - \tau_2;\sigma_1,\sigma_2) =
 \frac{i \eta_2}{2\pi\sqrt{- \xi_G^2}}
{\bf K}\left(\frac{1}{\xi_G^2}\right)\;,
\quad \xi_G^2 \in \C\backslash \R^+\,,
\label{Dglobal}
\ee
where $\eta_2 = {\rm sign}[\cos{\rm Re}(\frac{\tau_1 - \tau_2}{2})]$. 
It is $4\pi$-periodic in ${\rm Re}(\tau_1 -\tau_2)$, signaling that the theory lives on the 
double cover of ${\rm AdS}_2$. The result (\ref{Dglobal}) arises through analytic 
continuation from the mode sum $W(\tau_1 - \tau_2 - i\eps; \sigma_1,\sigma_2)= 
i \sum_{n \geq 0} \varphi_n(\tau_1 -i\eps,\sigma_1)\varphi_n(\tau_2,\sigma_2)^*$. 
The boundary value for real points comes out as 
\be
W(\tau_1 -\tau_2;\sigma_1,\sigma_2) = \eta_2 \frac{1}{2} 
\Big[{\rm sign}[\sin(\tau_1 - \tau_2)]\, d(\xi_G^2) + i d(1 - \xi_G^2)\Big]\;,
\label{DGreal}
\ee 
with $d$ given by (\ref{D3}). This agrees with (\ref{D1}) in the patch covered 
by the Poincar\'{e} coordinates. In addition (\ref{DGreal}) allows one to understand 
the close relation between the real and the imaginary part of (\ref{D1}).
From $\varphi_n(\tau_1 + \pi, -\sigma_1) = -i \varphi_n(\tau_1,\sigma_1)$
and the mode sum one expects that $W(\tau + \pi;-\sigma_1,\sigma_2) = 
-i W(\tau;\sigma_1,\sigma_2)$, which indeed is a property of (\ref{DGreal}). 
Finally (\ref{Dglobal}) also illustrates that the causal properties of a quantum 
field theory living on $\widehat{{\rm AdS}}_2$ are very different from one 
living on ${\rm AdS}_2$. For example the pair of points $(\tau,\sigma)$ and 
$(\tau + \pi,\sigma)$ has embedding coordinates 
$(X^0,X^1,X^2) = (\sin\tau,-\sin\sigma,\cos\tau)/\cos\sigma$ and $(-X^0,X^1,-X^2)$, 
respectively. In a quantum field theory on ${\rm AdS}_2$ proper such 
points are expected to be causally independent in the sense that the corresponding 
smeared field operators commute. In the theory at hand however the commutator 
function at the points considered, having $\xi^2_G \geq 1$, does not vanish.

For the Killing vectors ${\bf e},\,{\bf h},\,{\bf f}$ the use of global 
coordinates amounts to switching to a $su(1,1)$ basis $\he,\,\hh,\,\hf$ 
of the $sl_2$ Lie-algebra, where $\hh$ is hermitian and $\he^{\dagger} = 
-\hf$ wrt the measure $d\tau d \sigma/\cos^2\sigma$ on ${\rm AdS}_2$. 
Repeating the previous analysis leads to known Noether charges \cite{NY}
\be
\widehat{H} = \sum_{n \geq 0} (2 n+1) a_n^{\dagger} a_n\;,\quad 
\widehat{E} = \sum_{n \geq 0} (n+1) a_n^{\dagger} a_{n+1} \;,\quad 
\widehat{F} = -\sum_{n \geq 0} (n+1) a_{n+1}^{\dagger} a_n \;,
\label{su11charges}
\ee
where $\widehat{H}$ is hermitian and $\widehat{E}^{\dagger} = -\widehat{F}$.
In our conventions $[\widehat{H}, \Phi(\tau,\sigma)] = -\hh \Phi(\tau,\sigma)$,
$[\widehat{E}, \Phi(\tau,\sigma)] = \he \Phi(\tau,\sigma)$,
$[\widehat{F}, \Phi(\tau,\sigma)] = \hf \Phi(\tau,\sigma)$. 
The 1-mode Fock space $\bigoplus_n \C a_n^{\dagger} |0\ket$ carries a unitary 
irreducible representation of the $su(1,1)$ realization (\ref{su11charges}),
with $a_0^{\dagger}|0\ket \sim Q_{\pm} |0\ket$ playing the role of the 
lowest weight vector. 

In contrast to the $SL(2,\R)$ action in (\ref{SL2}) the finite transformations 
generated by the exponentials of $\he,\,\hh,\,\hf$ act geometrically on the 
covering space. On functions of the exponentials $e^{i(\tau \pm \sigma \pm \pi/2)}$ 
the $SU(1,1)$ action is given by
\be 
e^{i(\tau \pm \sigma \pm \pi/2)} \rra
\frac{\alpha \,e^{i(\tau \pm \sigma \pm \pi/2)} +\beta}%
{\beta^* e^{i(\tau \pm \sigma \pm \pi/2)} + \alpha^*}\;,\sspace
|\alpha|^2 - |\beta|^2 =1\;.
\label{SU11}
\ee
It is unitarily implemented by the charges (\ref{su11charges}) and 
yields a representation $SU(1,1) \ni g \ra U(g)$ under which 
the field $\Phi(\tau,\sigma)$ transforms covariantly with weight zero
\be
U(g) \Phi(\tau,\sigma) U(g)^{-1} = \Phi(\tau^g,\sigma^g) \;,
\sspace g = {\!\!\alpha \;\;\; \beta \choose \beta^* \;\; \alpha^* }
\in SU(1,1)\;,
\label{UPhi}
\ee
with $(\tau^g,\sigma^g)$ induced by (\ref{SU11}). As they act on the same 
Fock space, and in view of the isomorphism $SU(1,1) \simeq SL(2,R)$, we use the 
same notation $U(g)$ as in (\ref{SL2AFT}). The correspondence between the 
$SU(1,1)$ parameters in (\ref{SU11}) and the $SL(2,\R)$ parameters in (\ref{SL2}) 
is $\alpha = [a + d + i(b-c)]/2,\;\beta = [-a + d + i(b+c)]/2$.
For the original scalar field $\phi(\tau,\sigma) =
[(\cos\tau - \sin\sigma)/\cos\sigma]^{1/2}\Phi(\tau,\sigma)$
eq.~(\ref{UPhi}) implies the transformation law
\be
U(g) \phi(\tau,\sigma) U(g)^{-1} = \frac{1}{|\alpha + \beta^*|}
\left(\frac{\cos\tau - \sin\sigma}{\cos(\tau + 2\gamma) - \sin\sigma}\right)^{1/2}
\phi(\tau^g,\sigma^g) \;,\quad \gamma= {\rm Arg}(\alpha + \beta^*)\;.
\label{Uphi}
\ee
The $sl(2,\R)$ Noether charges now have globally defined counterparts 
\be
E_G = \frac{i}{2}[(\widehat{E} + \widehat{F}) -i \widehat{H}]\;,\quad
H_G = \widehat{E} - \widehat{F}\;,\quad
F_G = \frac{i}{2}[(\widehat{E} + \widehat{F}) +i \widehat{H}]\;,
\ee
which on the patch of $\widehat{{\rm AdS}}_2$ covered by the Poincar\'{e} coordinates 
generate the same geometric action as $E,\,H,\,F$.
They are of interest because of their physical meaning for the original problem:
$2 E_G$ is the $C$-energy $\gamma_{\infty}$, now for globally defined solutions,
and similarly for $H_G$ and $F_G$. The unitary representation $U(g)$ rotates 
these charges into each other
\be
U(g) \left( \begin{array}{c} E_G \\ H_G\\ F_G \end{array}\right) U(g)^{-1} = 
\left( \begin{array}{ccc} a^2 & - a c & - c^2 \\
- 2 a b & a d + bc & 2 c d \\
- b^2 & b d & d^2
\end{array} \right)
\left( \begin{array}{c} E_G \\ H_G\\ F_G \end{array}\right)\;,
\label{SL2charge} 
\ee
where we used the $SL(2,\R)$ parameterization.  
The rotation matrices are elements of $SO_0(2,1)$ in accordance with 
the isomorphism $SL(2,\R)/Z_2 \simeq SO_0(2,1)$. The classical analogue
of (\ref{SL2charge}) has the following interpretation:      
The $SL(2,R)\simeq SU(1,1)$ group action maps a solution $\phi(\tau,\sigma)$ 
of the field equation into a new solution, and in the global $(\tau,\sigma)$ 
coordinates it is a point transformation. One can thus ask whether the values 
of the conserved charges (energy content etc.) of the new solution can be 
expressed in terms of that of the old solution. 
The answer is Yes and is given by the rhs of (\ref{SL2charge}).  
In view of (\ref{ER3}) this implies for the classical spectra (i.e.~the
possible values on a solution): ${\rm cSpec}E_G = \R^+$,
${\rm cSpec}H_G = \R$, ${\rm cSpec}F_G = \R^-$. In the quantum theory 
the same holds, now with ${\rm cSpec}$ referring to the continuous 
spectrum of the corresponding operators. In addition the spectrum of 
$\widehat{H} = E_G - F_G$ is discrete and non-negative.

{\em CPT, wedges, and thermalization:} 
Geometrically the proper CPT conjugation on $\widehat{{\rm AdS}}_2$ is 
\be
j(\tau,\sigma) = (-\tau + \pi,-\sigma)\;,\sspace j^2 = id\;.
\label{CPT1}
\ee
Only for fields that are $2\pi$-periodic in $\tau$ would $j$ project down
to a CPT operation on ${\rm AdS}_2$ proper. Of course $\phi(\tau,\sigma)$
is $2\pi$ anti-periodic in $\tau$. For the ER-system the
following version of a CPT theorem holds: There exists a
conjugate-linear anti-unitary operator $J$ uniquely determined by the 
properties
\be 
J \Phi(\tau,\sigma)J^{-1} = \Phi(j(\tau,\sigma))^{\dagger}\;,
\sspace J |0\ket = |0\ket\;.
\label{CPT2}
\ee
We omit the proof. 
A point $(\tau,\sigma)$ and its conjugate $j(\tau,\sigma)$ are 
spacelike separated, $\xi^2_G < 0$, iff $ \cos 2\sigma + \cos 2 \tau <0$.
This partitions $\widehat{\rm AdS}_2$ into wedges $W_k,\,k \in \Z$, adjacent 
to the $\sigma = -\pi/2$ boundary, their CPT conjugates, and central 
diamond-shaped regions separating the wedges; c.f.~Fig.~2.

\begin{figure}[bht]
\hspace{4.5cm}
\leavevmode
\epsfxsize=67mm
\epsfysize=87mm
\epsfbox{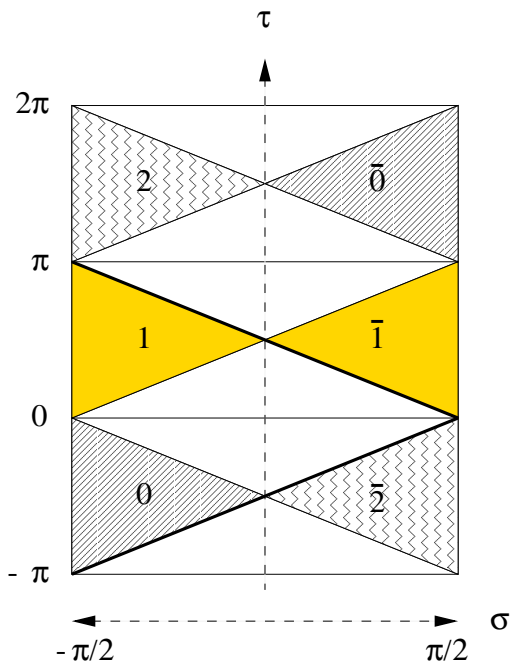}

\caption{Subspaces of $\widehat{\rm AdS}_2$. The patch covered by the
Poincar\'{e} coordinates $(t,r)$ is the bold bounded triangle; the wedges
$W_k$ and $W_{\bar{k}}$, $k=0,1,2,\ldots$, are CPT conjugate to each other.}
\end{figure}
                                                                 
There is a unique wedge $W_1$ which together with its CPT conjugate lies in the 
primitive domain. It is characterized by the condition $t \pm r = 
\tan\frac{1}{2}(\tau \pm \sigma \pm \pi/2) \geq 0$, and thus 
in Poincar\'{e} coordinates is the cone $t \geq  r \geq  0$. 
A convenient set of coordinates on $W_1$ is $(\xi,\rho),\,\xi \in \R,\,
\rho \geq 0$, related to $(t,r)$ by $t = e^{\xi} \ch \rho,\,r = e^{\xi} \sh \rho$. 
Denoting by $\Phi(\xi,\rho)$ the field $\Phi(t,r)|_{W_1}$ in these coordinates 
one can use (\ref{qSL2}) to identify $H/2$ as the Hamiltonian 
for the cone, i.e. 
\be
e^{is H} \Phi(\xi,\rho) e^{-i s H} = \Phi(\xi + 2 s, \rho)\,,
\quad s\in \R\;.
\label{wedge3}
\ee
Let us write $W(\xi_1 -\xi_2;\rho_1,\rho_2)$ for the two-point function
$W(t_1 -t_2; r_1,r_2)$ restricted to the cone $W_1$, where $(t_i,r_i)= 
(e^{\xi_i} \ch\rho_i, e^{\xi_i} \sh \rho_i)$, $i=1,2$. It enjoys the 
following Kubo-Martin-Schwinger property, which results in a thermalization 
phenomenon akin to the Unruh effect: As a function of $\xi_1 - \xi_2$ 
$W$ is analytic in the strip $- 2\pi < {\rm Im}(\xi_1 - \xi_2) < 0$. 
The boundary values at the lower and upper rim of the strip are 
related by 
\be
W(\xi_1 - \xi_2 - 2 i\pi + i \eps; \rho_1,\rho_2) = 
W(\xi_2 - \xi_1 - i\eps;\rho_2,\rho_1)\;. 
\label{KMS}
\ee 
The proof is based on the primitive tubes of analyticity in \cite{Bertetal} 
and the above CPT theorem; we omit the details. Let us emphasize that the result 
holds both for spacelike and timelike separated points $(t_1,r_1),\,(t_2,r_2) \in W_1$; 
it does not refer to the worldlines of ``observers'' but is a property of the 
cone $W_1$: Restricting the quantum theory of the ER-scalar to $t \geq r \geq 0$, 
the Fock vacuum is a thermal state for the dilatation-evolution (\ref{wedge3}), with 
temperature $1/2\pi$. From the 4D viewpoint the restricted theory can be regarded 
as the quantum theory of those ``exotic'' ER-waves whose scalar $\phi(t,r)$ 
has support in $W_1$ only. The latter can be constructed explicitly
by solving the wave eq.~(\ref{AdSwave}) in the $(\xi,\rho)$ coordinates.
The ground state for $H$ however is not $SL(2,\R)$ invariant \cite{Strom}.
In the context of the ER-system the above thermal representation of 
the canonical commutation relations is therefore preferred in that it preserves
more of the classical symmetries.   
\pagebreak

{\em Boundary holography:} 
In global coordinates the $r=0$ axis corresponds to the $\sigma = -\pi/2$
boundary of ${\rm AdS}_2$. Parallel to (\ref{reconstr1}) we wish to reconstruct 
the field $\phi(\tau,\sigma)$ from its $\sigma = -\pi/2$ boundary values. 
Since $\phi(\tau,\sigma)$ is $2\pi$ anti-periodic in $\tau$ one can restrict 
attention to the primitive domain $-\pi < \tau <\pi$. It is convenient to 
introduce the mode sums 
\be
\!\!
a^-(\th) = \sum_{n \geq 0} a_n (-)^n e^{-i (n+1/2)\th},\;\;
{\rm Im}\,\th < 0\;,\quad 
a^+(\th) = \sum_{n \geq 0} a^{\dagger}_n (-)^n e^{i (n+1/2)\th},
\;\;{\rm Im}\,\th > 0\,,
\ee
both declared to vanish in the other half-plane. They are $SU(1,1)$ densities 
of weight $1/2$, i.e.
\be
U(g) a^{\pm}(\th) U(g)^{-1} = 
\frac{a^{\pm}(\th^g)}{|\beta e^{i\th} + \alpha|}\;. 
\label{Ua}
\ee
In terms of them we define the boundary field $\phi^b(\tau)$ by
\be
\phi^b(\tau) = \cos\frac{\tau}{2}
\lim_{\eps \ra 0}\, [a^+(\tau + i\eps) + a^-(\tau -i\eps)]\;.
\ee
Classically the $a^{\pm}(\th)$ play the role 
of scattering data wrt the $\tau$ evolution. One can recover them from a
given $\phi^b(\tau)$ by factorizing it into a sum of functions $\cos\frac{\th}{2}
a^{\pm}(\th)$ holomorphic in the upper and the lower half plane, respectively. 
The factorization has a unique solution in the primitive domain given by 
\be
\frac{1}{2 \pi}\int_{-\pi}^{\pi} d\tau 
\frac{\phi^b(\tau)}{1-e^{-i(\tau -\th -i\eps)}} =
\cos\frac{\th}{2}\, a^-(\th - i\eps) + \frac{1}{2} a_0^{\dagger}\;,
\ee
together with its hermitian conjugate yielding $a^+(\th)$. The zero modes are 
easily split off and the classical reconstruction is completed by insertion 
of $a^{\pm}(\th)$ into 
\begin{subeqnarray}
\phi(\tau,\sigma) &=& \frac{1}{2\pi} \int_{-\pi}^{\pi} d\th\, 
[a^-(\th)\, \phi_{\th}(\tau,\sigma) + a^+(\th)\, \phi_{\th}(\tau,\sigma)^*]\;,
\\
\phi_{\th}(\tau,\sigma) &=& \frac{1}{\sqrt{2}}
\frac{(\cos \tau - \sin \sigma)^{1/2} \;e^{i(\th - \tau)/2}}%
{[1 + 2 \sin\sigma e^{i(\th -\tau + i \eps)} + 
e^{2 i (\th - \tau + i\eps)}]^{1/2}}\;,\quad \eps > 0\;,
\label{holo1}
\end{subeqnarray}
where $-\pi < \tau,\th < \pi$. Since all transformations are linear the above 
procedure carries over to the quantum case when recast in terms of generic matrix 
elements of $\phi$. The $SU(1,1)$-covariance of the data $\phi^b(\tau)$, 
i.e.~(\ref{Uphi}) at $\sigma = -\pi/2$,  implies (\ref{Uphi}) for the 
reconstructed field operator. 


\end{document}